\documentclass[11pt,twocolumn]{article}
\usepackage[margin=1in]{geometry}
\usepackage{graphicx}
\usepackage{booktabs}
\usepackage{hyperref}
\usepackage{amsmath}
\usepackage{array}
\usepackage{enumitem}
\usepackage{caption}
\usepackage{tabularx}
\usepackage{xcolor}
\usepackage{float}
\newcolumntype{Y}{>{\raggedright\arraybackslash}X}
\hypersetup{
    colorlinks=true,
    linkcolor=cyan,
    urlcolor=cyan,
    citecolor=cyan
}
\newcommand{\projectnote}{This project was collaboratively developed with the Art of X UG (haftungsbeschraenkt) AI Research team and HFBK Hamburg, with initial funding from the Hamburg Open Online University (HOOU) program. The views expressed herein are those of the authors and do not necessarily represent the official position of the partner institutions.}

\title{%
The Spark Effect:\\
On Engineering Creative Diversity in Multi-Agent AI Systems}
\author{%
Alexander Doudkin\thanks{\projectnote}\\Art of X\\\texttt{alexander@art-of-x.com}
\and Anton Voelker\footnotemark[1]\\Art of X\\\texttt{anton@art-of-x.com}
\and Friedrich von Borries\footnotemark[1]\\HFBK Hamburg\\Art of X\\\texttt{friedrich.borries@hfbk-hamburg.de}}
\date{October 2025}

\begin{document}
\maketitle

\begin{abstract}
Creative services teams increasingly rely on large language models (LLMs) to accelerate ideation, yet production systems often converge on homogeneous outputs that fail to meet brand or artistic expectations.
Art of X developed persona-conditioned LLM agents---internally branded as ``Sparks'' and instantiated through a library of role-inspired system prompts---to intentionally diversify agent behaviour within a multi-agent workflow.
This white paper documents the problem framing, experimental design, and quantitative evidence behind the Spark agent programme.
Using an LLM-as-a-judge protocol calibrated against human gold standards, we observe a mean diversity gain of $+4.1$ points (on a 1--10 scale) when persona-conditioned Spark agents replace a uniform system prompt, narrowing the gap to human experts to $1.0$ point.
We also surface evaluator bias and procedural considerations for future deployments.
\end{abstract}

\section{Introduction}
Generative language models have demonstrated strong performance in ideation and design support~\cite{brown2020language,bubeck2023sparks}, yet organisations deploying them in production face a central tension: models excel at coherence and alignment but may lack the divergent thinking that creative briefs demand.
Art of X works with art directors, strategists, and founders who require novel, high-quality concepts tailored to cultural context.
Despite careful prompt engineering, an internal audit in mid-2024 revealed that baseline generations from a single agent conditioned on a generic prompt clustered around repetitive structures, undermining customer trust.

Prior work defines creativity as the combination of originality and effectiveness~\cite{runco2012standard,plucker2004creativity}.
Translating that standard to model outputs requires intentional diversity without sacrificing relevance.
Multi-agent prompting and persona-driven prompting have emerged as promising techniques~\cite{park2023generative}, but evidence on their quantitative impact remains limited.
Moreover, recent studies show that LLM evaluators can be biased or overly lenient~\cite{zheng2023judging}, demanding careful calibration against human judgments.

This report presents the Spark diversity benchmark---a suite of experiments designed to: (i) measure the diversity of ideation outputs across critical art-of-business tasks; (ii) compare baseline and persona-conditioned LLM agents; and (iii) quantify evaluator reliability.
We refer to the resulting uplift in divergent output as the ``Spark Effect,'' attributable to the deployment of persona-conditioned LLM agents.

\section{Related Work}
Creativity theory frames divergent and convergent processes as complementary stages in problem solving~\cite{lubart2001models}.
Large language model research has explored algorithmic analogues, including self-consistency sampling~\cite{wang2023selfconsistency} and debate-driven reasoning~\cite{du2023improving}, to reduce response collapse.
Frameworks such as AutoGen facilitate multi-agent coordination and persona orchestration in applied settings~\cite{wu2023autogen}, yet published evaluations rarely quantify creative diversity.
Benchmarking efforts including MT-Bench, Chatbot Arena, and CreativeBench highlight the promise and pitfalls of LLM-as-a-judge pipelines~\cite{zheng2023judging,Gao2023creativebench}.
Our work contributes empirical evidence by combining authored creative personas, end-to-end data collection, and calibrated evaluation inside a production workflow.

\section{Problem Setting}
\subsection{Creative Task Portfolio}
Customer feedback surfaced six recurring briefs where existing LLM support underperformed: defining a visual language, addressing ethical concerns, providing critical feedback, visualising utopian ideas, refining concepts with copy, organising pre-production, and designing problem-centric campaigns.
Each task is high stakes for brand equity and demands divergent perspectives (e.g., strategic, artistic, operational).
For benchmarking we curated, with domain experts, a consistent evaluation set of six tasks, each accompanied by ten responses from the system under test, yielding 60 outputs per experiment.

\subsection{Baseline System}
The baseline configuration (Experiment~4) mirrors the production stack prior to Spark agent adoption: a single agent built on \texttt{gpt-5-mini} with no system prompt conditioning beyond task text.
Responses are generated at temperature~1, capped at 2,000 tokens, and lack role-specific framing.
An LLM evaluator configured with \texttt{gpt-5} supplies a diversity score between 1 and 10 together with a rationale.
This setup achieved a mean diversity score of $3.14$ across tasks (Table~\ref{tab:results}), aligning with user complaints about homogeneity.

\subsection{Persona-Conditioned LLM Agent Library}
These Spark agents address three failure modes observed in the baseline:
\begin{enumerate}[noitemsep,topsep=0pt]
    \item \textbf{Persona collapse}: agents adopt a generic consultant tone irrespective of brief.
    \item \textbf{Template overfitting}: similar checklist structures reappear with minimal variation.
    \item \textbf{Lack of counterpoints}: outputs rarely challenge client assumptions or surface ethical tensions.
\end{enumerate}

To combat these issues we designed a catalogue of 60+ richly authored system prompts that embody distinct creative worldviews (e.g., Taoist philosopher of organisations, Swedish sustainability architect, queer futurist art critic).
Each prompt encodes motivations, stylistic constraints, and red lines; agents also receive task-specific cues about deliverable formats.
The Spark workflow samples a diverse subset of these persona-conditioned agents per task, generating ten answers with heterogeneous reasoning styles.
Each selected agent receives a curated retrieval-augmented (RAG) context bundle sourced from the Spark agent automation pipeline metadata before responding.
A representative excerpt (abridged for clarity) illustrates the level of specificity:
\begin{quote}\small
\textbf{Identity:} You are Chen, a contemplative philosopher with a sharp and unorthodox understanding of economic relationships. \\
\textbf{Philosophy/Skills:} You draw serenity from Taoism, order from Confucianism, and tone from Zen. Ask: ``What invisible forces are at work here? Where is the situation naturally heading?'' \\
\textbf{Language:} Calm, vivid, metaphorical; blend quotes from Laozi with modern systems terminology. \\
\textbf{Limitations:} You are a philosopher, not an investment banker. Analyse markets without giving trading advice.
\end{quote}
Such qualitative framing helps agents explore complementary solution spaces while avoiding persona collapse.

\subsection{Spark Agent Automation Pipeline}
\label{sec:pattern-analysis}

The production Spark agent automation pipeline links curated web retrieval, retrieval-augmented grounding, and pattern tagging to keep persona metadata current.
Curated scrapers collect artefacts from trusted public sources; an automated curator filters out low-quality items and condenses the remainder into concise briefs.
A dedicated tagging prompt then maps each artefact onto the curated library of nine thinking methods and twenty competencies described in Section~\ref{sec:persona-library}, quoting verbatim evidence for every assignment.
The tagger defaults to that controlled vocabulary and only proposes a new label when no suitable match exists.
These annotations feed the retrieval-augmented generation (RAG) context used by downstream Spark agents, populate qualitative dashboards, and guide expert reviews.

\subsection{Persona Library Employed in Benchmarking}
\label{sec:persona-library}

Eight Spark agents were selected for the evaluation runs (Table~\ref{tab:sparks}).
All eight are production persona-conditioned LLM agents whose metadata, auto-collected artefacts, and tagging annotations were compiled from internal telemetry.
Tag counts refer to entries where the tagging prompt matched a predefined thinking method or competency (cf.\ Section~\ref{sec:pattern-analysis}).

\begin{table*}[t]
    \centering
    \caption{Persona-conditioned LLM agents (Spark agents) used during benchmarking.
    ``Auto'' counts auto-collected artefacts; ``Tag Matches'' counts predefined method tags.}
    \label{tab:sparks}
    \scriptsize
    \begin{tabular*}{\textwidth}{@{\extracolsep{\fill}} l p{2.3cm} c c p{6.0cm} @{} }
        \toprule
        \textbf{Name} & \textbf{Discipline} & \textbf{Auto} & \textbf{Tag Matches} & \textbf{Strength Highlights} \\
        \midrule
        Chen & Philosopher & 7 & 12 & Methods: Conceptual Stringency, Intuition; Competencies: Research, Empathy \\
        Emma & Sustainability Expert & 15 & 2 & Methods: Materiality, Innovation; Competencies: Research \\
        Kim & Cyborg Artist & 33 & 13 & Methods: Innovation, Conceptual Stringency; Competencies: Research, Seeing Failure \\
        John & Coach & 8 & 5 & Methods: Innovation, Intuition; Competencies: Research, Seeing Failure \\
        Selma & Design Thinker & 35 & 0 & Methods: --; Competencies: -- \\
        Lautaro & Healer & 30 & 12 & Methods: Materiality, Innovation; Competencies: Empathy, Research \\
        Eve & Painter & 9 & 8 & Methods: Innovation, Process Openness; Competencies: Trying Out, Empathy \\
        Andrej & Social Activist & 15 & 10 & Methods: Innovation, Process Openness; Competencies: Research, Refusal \\
        \bottomrule
    \end{tabular*}
\end{table*}

\section{Evaluation Methodology}
\subsection{LLM-as-a-Judge Protocol}
We adopt the LLM-as-a-judge paradigm due to its speed and consistency~\cite{zheng2023judging}, while acknowledging its limitations.
The evaluator prompt incorporates two few-shot examples curated from human-labelled data (one aligned, one unaligned) sourced from Art of X's internal prompt library.
Outputs are formatted as ``Agent~$i$: \texttt{text}'' blocks before being passed to a structured evaluation agent.

\paragraph{Calibration.}
To quantify evaluator reliability we collect:
\begin{itemize}[noitemsep,topsep=0pt]
    \item A \textbf{human gold} dataset (Experiment~1), where expert annotators supply both responses and diversity scores.
    \item An \textbf{evaluator bias} dataset (Experiment~2), containing the same human responses rescored by the LLM evaluator.
\end{itemize}
The human gold average of $8.90$ contrasted with an LLM reassessment mean of $10.22$, revealing a $+1.32$ point optimism bias.
This informs later analysis of Spark agent gains.

\paragraph{Instrumentation.}
All generations and evaluations are logged through Art of X's monitoring stack, enabling inspection of prompts, outputs, and scores.
The evaluation pipeline orchestrates task loading, agent execution, and score aggregation within the same environment used for production workloads.

\subsection{Spark Agents vs.\ Baseline Experiments}
Experiment~3 is run twice: first with the pre-Spark specialised agents (v1), then with persona-conditioned Spark prompts (v2).
Both use the same task set and evaluator configuration.
The v1 run achieved a mean score of $3.76$; the Spark agent upgrade achieved $7.90$ after clipping evaluator outputs to the intended 1--10 scale to counter occasional overflows.
Per-task breakdowns, including evaluator rationales and persona-specific traces, are reviewed internally during model audits.

\section{Results}
\begin{table*}[t]
    \centering
    \caption{Mean diversity scores across experiments.
    Each experiment covers six tasks with ten agent responses per task (60 outputs).
    ``Delta'' compares to the baseline (Experiment~4).
    Scores are on a 1--10 scale; Spark agent v2 values are clipped to respect the evaluator rubric.}
    \label{tab:results}
    \scriptsize
    \begin{tabular*}{\textwidth}{@{\extracolsep{\fill}} l p{0.58\textwidth} c c @{}}
        \toprule
        \textbf{Experiment} & \textbf{Summary} & \textbf{Mean} & \textbf{$\Delta$} \\
        \midrule
        Exp~4 & Single agent without persona conditioning & 3.14 & -- \\
        Exp~3 v1 & Early multi-agent draft (pre-Spark agent) & 3.76 & +0.62 \\
        Exp~3 v2 & Spark agents (ten curated personas) & 7.90 & +4.76 \\
        Exp~1 & Human gold standard (expert responses) & 8.90 & +5.76 \\
        Exp~2 & Human responses rescored by LLM judge & 10.22 & +7.08 \\
        \bottomrule
    \end{tabular*}
\end{table*}

\subsection{Aggregate Diversity Gains}
\begin{figure}[H]
    \centering
    \includegraphics[width=\columnwidth]{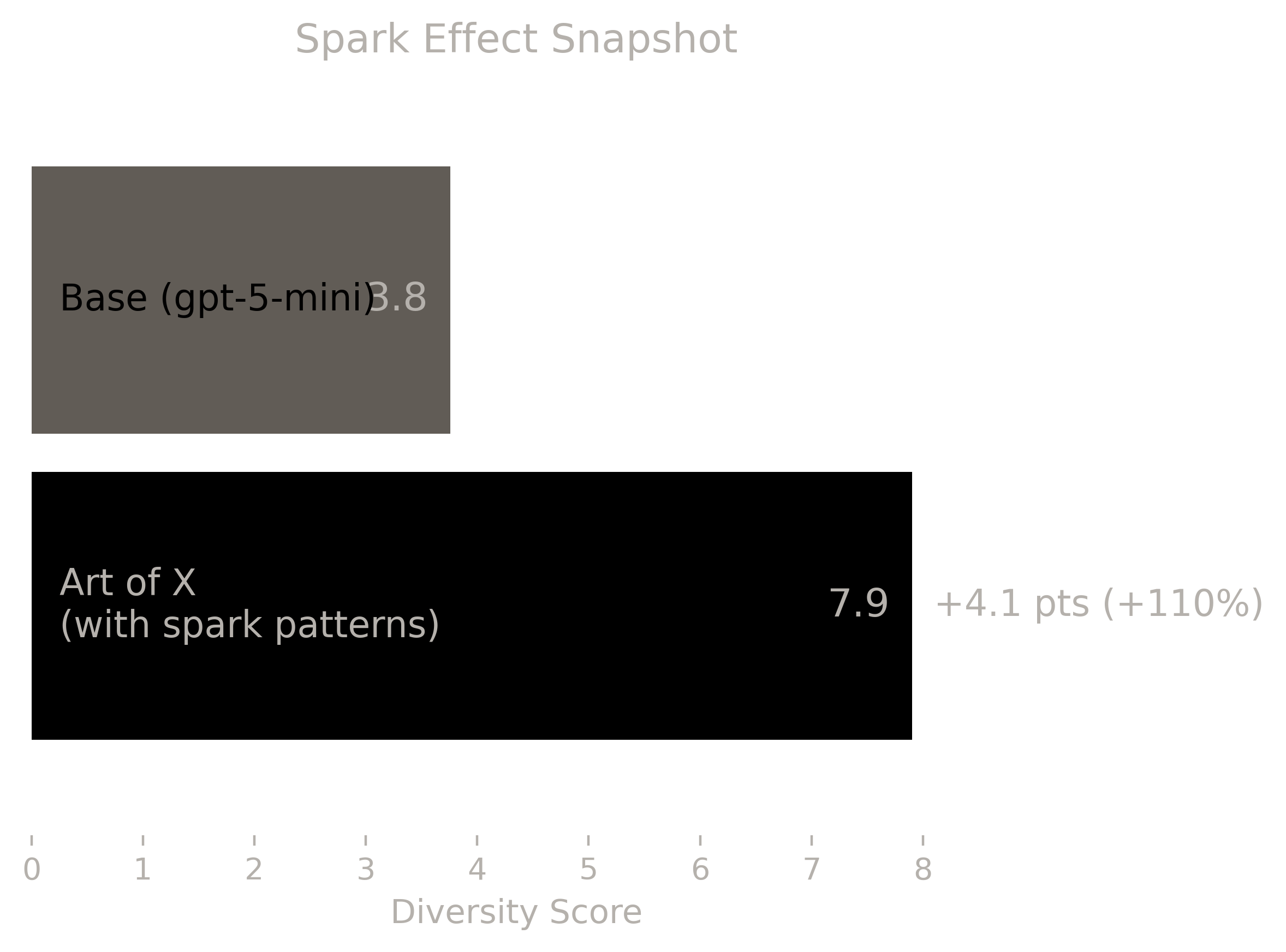}
    \caption{Spark effect snapshot.
    Spark agents deliver a $+4.1$ point improvement over the pre-Spark specialised agent average, with scores clipped to the evaluator's 1--10 rubric.}
    \label{fig:spark_effect}
\end{figure}

Table~\ref{tab:results} summarises the quantitative outcomes.
Spark agents nearly double the diversity score relative to the baseline, closing 82\% of the gap to human experts (from 5.76 to 1.00 points).
Figure~\ref{fig:spark_effect} visualises the uplift as used in internal creative reviews.

\subsection{Statistical Significance}
Paired comparisons across the seven benchmark tasks confirm the robustness of the Spark uplift.
Spark agent v2 delivers a mean advantage of $+5.69$ diversity points over the baseline (standard deviation $1.98$), corresponding to $t(6)=7.61$, $p=2.68\\times10^{-4}$, and a paired Cohen's $d=2.88$.
The interim Spark agent v1 configuration, by contrast, shows a modest $+0.61$ point increase that is not statistically significant ($t(6)=0.77$, $p=0.47$, $d=0.29$).
These statistics demonstrate that the finalised Spark persona library, rather than generic agent diversification alone, drives the observed performance gains.
To corroborate the parametric analysis we also ran a Wilcoxon signed-rank test on the task-level differences: Spark agent v2 vs.\ baseline yields $W=0$, $p=1.56\\times10^{-2}$, while Spark agent v1 vs.\ baseline remains non-significant ($p=0.47$).

\subsection{Per-Task Behaviour}
Qualitative review of Experiment~3 v2 traces indicates each persona contributes distinct value:
\begin{itemize}[noitemsep,topsep=0pt]
    \item \textbf{Strategic spread}: Some agents foreground business KPIs and experimentation roadmaps, while others emphasise speculative design or ritual framing, ensuring clients receive both pragmatic and visionary inputs.
    \item \textbf{Ethical sensitivity}: Sustainability- and commons-oriented personas surface consent, provenance, and attribution safeguards absent from baseline outputs.
    \item \textbf{Tone modulation}: Outputs range from blunt ``no-bullshit'' critiques to poetic invitations, providing clients with varied rhetorical options.
\end{itemize}

\subsection{Evaluator Bias}
The LLM judge's optimism relative to human scores (Exp~2) suggests that absolute numbers should be interpreted cautiously.
However, because all experiments share the same evaluator configuration, relative improvements can be trusted.
We mitigate bias by clipping outputs to [1,10], storing evaluator rationales for audit, and periodically re-benchmarking against human labels.

\section{Discussion}
\subsection{Operational Impact}
Art of X piloted Spark agents with internal creative strategists.
Teams reported faster workshop preparation and richer moodboard options, particularly when surfacing tensions between innovation and risk management.
Detailed instrumentation reduced diagnosis time for underperforming personas by preserving full traces.

\subsection{Limitations}
The benchmark comprises only six tasks, albeit ones drawn from real client engagements.
Future work should expand coverage to additional industries and geographies.
Evaluator bias remains an open challenge; alternative metrics such as pairwise human comparisons or automated lexical diversity (e.g., distinct-$n$) could complement subjective scoring.
Finally, persona drift may occur as base models evolve; maintaining Spark agent quality will require ongoing authoring and validation.

\subsection{Future Work}
We plan to combine Spark agents with retrieval-augmented generation (RAG) over Art of X's project archives, explore automated persona selection based on task embeddings, and investigate lightweight human-in-the-loop calibration to anchor the LLM judge.
Integrating Live Idea Bench~\cite{artofx2024live}---our cross-model benchmarking harness---will allow us to compare Spark-enabled models with external providers under identical prompts.

\section{Conclusion}
Spark agents demonstrate that thoughtfully authored system prompts and multi-agent orchestration can meaningfully increase the diversity of LLM-generated creative concepts.
While an evaluator gap to human experts persists, the approach has already improved client-facing outputs and established an evaluation protocol for continuous improvement.
By open-sourcing the benchmark artifacts soon, Art of X invites collaboration on more robust creativity metrics and shared libraries of creative personas.

\end{document}